\documentclass[aps,pra,superscriptaddress,preprintnumbers,twocolumn,showpacs]{revtex4}
\usepackage{amsmath}
\usepackage{amssymb}
\usepackage{graphicx}
\usepackage{color}
\usepackage{bm}

\newcommand{\e}{\text{e}}

\renewcommand{\Re}{\mathop{\text{Re}}\nolimits}
\renewcommand{\Im}{\mathop{\text{Im}}\nolimits}

\newcommand{\ket}[1]{|{#1}\rangle}
\newcommand{\bra}[1]{\langle{#1}|}

\newcommand{\bracket}[2]{\langle#1|#2\rangle}

\def\bbm[#1]{\mbox{\boldmath$#1$}}

\definecolor{dgreen}{rgb}{0,0.5,0}

\definecolor{delete}{cmyk}{0.5,0,0,0}
\definecolor{deletey}{cmyk}{0,0.5,0,0}

\begin{document}
\title{Extraction of a squeezed state in a field mode via repeated measurements on an auxiliary quantum particle}
\author{Bruno Bellomo}
\author{Giuseppe Compagno}
\affiliation{CNISM \& Dipartimento di Scienze Fisiche ed Astronomiche,
Universit\`{a} di Palermo, via Archirafi 36, 90123 Palermo, Italy}
\author{Hiromichi Nakazato}
\affiliation{Department of  Physics, Waseda University, Tokyo 169-8555, Japan}
\author{Kazuya Yuasa}
\affiliation{Waseda Institute for Advanced Study, Waseda University, Tokyo 169-8050, Japan}
\begin{abstract}
The dynamics of a system, consisting of a particle
initially in a Gaussian state interacting with a field mode,
under the action of repeated measurements performed on the
particle, is examined. It is shown that regardless of its initial state the field is distilled into a squeezed state.
The dependence on the physical parameters of the dynamics is investigated.
\end{abstract}

\pacs{03.65.Xp, 42.50.Dv}

\maketitle

\section{Introduction}\label{intro}
Obtaining pure states through distillation processes is a key step
to initialize and to control quantum systems in the field of quantum technology \cite{Nielsen-Chuanglibro2000,ref:BouwmeesterZeilingerEkert}.
For instance, state preparation is fundamental in several algorithms in quantum information and computation. So far various protocols for distillation have been proposed \cite{ref:BouwmeesterZeilingerEkert,Bennett1996,ref:Bennett1996PRA,Cirac1999}.

A procedure for distillation through Zeno-like measurements has been presented in \cite{Nakazato2003,Nakazato2004}. The protocol there considered regards a generic bipartite system made of two interacting subsystems $P$ and $F$.
The unitary dynamics of the total system is interrupted at regular intervals by measurements performed on one of the two subsystems, say $P$. These measurements affect strongly the dynamics of the non-measured system $F$, which is then governed by some effective evolution operator.
It is shown that if the spectrum of this operator satisfies certain conditions, then the non-measured system itself is driven toward a pure state irrespective of its initial conditions. This final state depends on the parameters of the total Hamiltonian, the measurement one performs, i.e., the state on which the system is projected by the measurement, and the time interval between measurements.
In this distillation protocol measurements are performed repeatedly as in the case of quantum Zeno effect \cite{sudarshan1977,nakazato1996,ref:QZE-review-HomeWhitaker,ref:QZE-review-Koshino,ref:QZE-review-PaoloSaverio} but here the interval between measurements is kept small but finite considering the limit of large number of measurements.

This general procedure can be useful to initialize quantum systems and it has been also extended to the case where the system is made of many parts showing that it is possible to distillate entangled states \cite{Nakazato2004,Wu2004}. For example it has been shown that it is possible to establish entanglement between two separate systems via repeated measurements on a third entanglement mediator \cite{Compagno2004}.
The effect of environmental noise on the protocol has been also investigated, showing that in the case of only dephasing effects one can still extract pure states \cite{Militello2007}.

However, these protocols have been analyzed only when the measured system has a discrete spectrum. Because of the conditions required for distillation, it is not obvious that distillation can be obtained even when the measured system has a continuous spectrum.
The main purpose of this paper is thus to apply the general procedure, as considered in \cite{Nakazato2003}, to the case where the measured system has a continuous spectrum. The system considered is a particle (system $P$), characterized by a continuous spectrum, interacting with a field mode (system $F$).
Our protocol consists of measurements performed on the particle $P$ to confirm it to be in a Gaussian state. We will show that the field mode is driven toward a pure state and we will examine how this distillation process depends
on parameters like the mode frequency $\omega$, the frequency of measurements
$\tau^{-1}$, the coupling between the particle and the field mode $g$, and
the characteristics of the Gaussian state of the particle $P$ to be confirmed repeatedly.

The paper is organized as follows. In Sec.\ \ref{par:Description of
the system} we describe the features of the model we consider. In
Sec.\ \ref{par:Description of the protocol} we describe
the protocol and in Sec.\ \ref{par:Application of
the protocol} we obtain and diagonalize the projected evolution operator. In
Sec.\ \ref{par:Distillation limit} we discuss distillation in terms of physical parameters of the system. In Sec.\ \ref{par:Conclusion} we summarize and
discuss our results. In Appendices \ref{app:Time-evolution operator} and \ref{app:Choise of sign} we collect some of the calculations, which are omitted in the text in order to keep the readability.

\vspace*{-0.1in}

\section{Model}\label{par:Description of the system}
We consider a particle of mass $m$ interacting with a single field mode of
frequency $\omega$. The particle interacts linearly with the field mode and the Hamiltonian reads as
\begin{equation}\label{hamiltoniana di partenza}
  \hat{H}=\frac{\hat{p}^2}{2m}+\hbar \omega
 \left( \hat{a}^{\dag}\hat{a}+\frac{1}{2}\right)+
   g \hat{p}(
  \hat{a}^{\dag}
  +\hat{a}
),
\end{equation}
where $\hat{p}$ is the particle momentum operator,
$\hat{a}$ and $\hat{a}^{\dag}$ are the
annihilation and creation operators of the field mode satisfying
the commutation rules,
$[\hat{a},\hat{a}^{\dag}]=1$, etc., and the real parameter $g$ is the
coupling constant.
This Hamiltonian is analogous to the electromagnetic Hamiltonian
restricted to the single-mode case and in dipole approximation
\cite{Petruccione-Breuerlibro2002,Bellomo2006}.

In the interaction picture the interaction Hamiltonian of
Eq.\ (\ref{hamiltoniana di partenza}) at time $s$ takes the form
\begin{equation}\label{hamiltoniana}
   \hat{H}^I(s)= g \hat{p}
  (
  \hat{a}^{\dag}\mathrm{e}^{i \omega s}
  + \hat{a}\,  \mathrm{e}^{-i \omega s}) .
\end{equation}
The evolution determined by this Hamiltonian can be treated exactly since its
commutator at two different times,
\begin{equation}\label{commutatore}
  [\hat{H}^I(s'),\hat{H}^I(s'')]=-2ig^2\hat{p}^2 \sin\omega
  (s'-s''),
\end{equation}
commutes with the Hamiltonian itself. This allows one to obtain the
exact evolution operator \cite{Petruccione-Breuerlibro2002,palma-suominen-ekert1996} as
\begin{align}\label{operatore evoluzione}
  &\hat{U}^I(\tau)
  =\mathrm{T}_{\leftarrow}\exp\! \left[
  -\frac{i}{\hbar}\int_0^\tau
  \mathrm{d}s\,
  \hat{H}^I(s)
  \right] \nonumber \\
&\quad
  = \exp\! \left[ -\frac{1}{2\hbar^2}\int_0^\tau
  \mathrm{d}s  \int_0^\tau
  \mathrm{d}s'\,\theta (s-s')[\hat{H}^I(s),\hat{H}^I(s')]\right]
  \nonumber \\
&\qquad{} \times \exp\! \left[
  -\frac{i}{\hbar}\int_0^\tau \mathrm{d}s\,
  \hat{H}^I(s)\right],
\end{align}
where $T_{\leftarrow}$ is the time ordering operator and $\theta
(s-s')$ is the Heaviside step function. Using
Eqs.\ (\ref{hamiltoniana})--(\ref{commutatore}) and
\begin{equation} \label{integrale su tempi}
  \int_0^\tau \mathrm{d}s \int_0^\tau
  \mathrm{d}s'\,\theta (s-s') \sin\omega(s-s')
  =\frac{\omega \tau-\sin\omega \tau}{\omega^2
  },
\end{equation}
the time evolution operator at time $\tau$ can be put in the
form
\begin{align} \label{operatore evoluzione cbh}
  \hat{U}^{I}(\tau)={} & \exp\!\left[i   g^2\hat{p}^2
   \frac{ \omega \tau-  \sin\omega \tau}
  {\hbar^2\omega^2}\right]
  \nonumber \\ & \times \exp\!\left[  g \hat{p}\left(
  \hat{a}^{\dag} \frac{1-\mathrm{e}^{i\omega \tau}}{\hbar \omega}-\hat{a} \frac{1-\mathrm{e}^{-i\omega \tau}}{\hbar \omega}\right)
  \right].
\end{align}
Indicating with $\hat{U}_0$ the time evolution operator associated to the
free part of the Hamiltonian of Eq.\ (\ref{hamiltoniana di
partenza}), the time evolution operator in the Schr\"{o}dinger picture, $\hat{U}(\tau)= \hat{U}_0(\tau)\hat{U}^{I}(\tau) $, is
given by
\begin{align}\label{evolution in S picture}
   \hat{U}(\tau)={} &\exp\!
\left[-\frac{i}{\hbar}\frac{\hat{p}^2 \tau}{2 m}  \left(1 -\frac{2 mg^2
 }{ \hbar\omega }
   \frac{ \omega \tau-
  \sin \omega \tau}{\omega \tau}\right) \right]
  \nonumber \\ &{} \times\exp\!
\left[-i \omega \tau
 \left( \hat{a}^{\dag}\hat{a}+\frac{1}{2}\right)\right]\nonumber \\ & {}\times  \exp\! \left[  g\hat{p} \left(
  \hat{a}^{\dag} \frac{1-\mathrm{e}^{i\omega \tau}}{\hbar \omega}-\hat{a} \frac{1-\mathrm{e}^{-i\omega \tau}}{\hbar \omega}\right)\right].
\end{align}

\vspace*{-0.1in}

\section{Protocol} \label{par:Description of the protocol}
In this section we briefly review the distillation procedure based on repeated measurements \cite{Nakazato2003} as applied to our system.

Assume that at time $t=0$ the particle is prepared in a pure state
$|\Phi_0\rangle$ and the field in an arbitrary mixed state,
denoted by $\rho_F(0)$. The unitary dynamics of the total system, governed by the time-evolution operator $\hat{U}(\tau)$ in Eq.\ (\ref{evolution in S picture}), is interrupted by the measurements performed
on the particle at intervals $\tau$.
Each time, the measurement projects the particle in its initial state $\ket{\Phi_0}$.
This action is represented by the projection operator $\mathcal{O} =
|\Phi_{0}\rangle \langle \Phi_{0} |\otimes \mathbf{1}_F $.  The
projection is partial on the total system because only the state of the particle is
set back to its initial state while the field is not initialized,
even though its dynamics is certainly affected by the
measurements. The total system after $N$ measurements is described by
\begin{equation}\label{total density matrix evolution}
     \hat{\rho}_T^{\tau}(N)\propto [\mathcal{O} \hat{U}(\tau)]^N[|\Phi_{0}\rangle \langle \Phi_{0}|\otimes   \hat{\rho}_F(0)]
      [ \hat{U}^{\dagger}(\tau) \mathcal{O}]^N .
\end{equation}
Following \cite{Nakazato2003}, we introduce the projected evolution operator between two
consecutive measurements,
\begin{equation}\label{contracted evolution operator}
    \hat{V}_{\tau}= \langle \Phi_{0}  | \hat{U}(\tau) |\Phi_{0}\rangle,
\end{equation}
so that, after \textit{N} measurements on the particle,
the field is described by the density matrix
\begin{equation}\label{field density matrix evolution}
     \hat{\rho}_F^{\tau}(N)=\frac{\hat{V}_{\tau}^N\hat{\rho}_F(0)
      \hat{V}_\tau^{\dagger N}}{P_{\tau}(N)}.
\end{equation}
Note that we retain only events in which the particle $P$ is found in the state $\ket{\Phi_0}$ by every measurement.
The normalization factor
\begin{equation}\label{probability}
    P_{\tau}(N)=\mathop{\mathrm{Tr}}\nolimits_F\{\hat{V}_{\tau}^N\hat{\rho}_F(0)
      \hat{V}_\tau^{\dagger N} \}
\end{equation}
represents such a probability (up to $N$ measurements) and gives the probability to obtain the state (\ref{field density matrix evolution}).

In this paper, we consider the following protocol: the particle $P$ is initially prepared in a Gaussian state $\ket{\Phi_0}$ and repeatedly projected on it at intervals $\tau$.
Such a measurement would be realized by switching on a harmonic potential, whose ground state coincides with the Gaussian state $\ket{\Phi_0}$, and seeing whether $P$ is in the ground state.	
The Gaussian state $\ket{\Phi_0}$ is characterized by the variances of the coordinate $\Delta r_{0}$ and of the momentum $\Delta p_{0}$, satisfying $\Delta r_{0}\,\Delta p_{0}=\hbar
/2$. In the momentum space, it is given by
\begin{equation}\label{initial state}
    |\Phi_{0} \rangle =\int \mathrm{d} p\,\frac{1}{\sqrt[4]{2 \pi(\Delta p_{0})^2}}\exp\!\left[-\frac{p^2}{4(\Delta p_{0})^2}\right]|p \rangle  ,
\end{equation}
where $|p \rangle $ are the eigenstates of the momentum operator $\hat{p}$, the initial
average momentum is $p_{0}=0$, and the initial average position
$r_{0}=0$.

Once chosen the measurements to perform on the particle, the next step is to ask if the protocol considered leads to purification of the field mode. In order to achieve purification for the non-measured system, the spectrum of the projected evolution operator $\hat{V}_{\tau}$ defined in Eq.\ (\ref{contracted evolution operator}) must satisfy the following conditions \cite{Nakazato2003}: its largest (in magnitude) eigenvalue must be unique, discrete, and nondegenerate.
In the next section we shall compute the projected evolution operator $\hat{V}_{\tau}$ for the present setup and then diagonalize it in order to analyze its spectrum.

\vspace*{-0.1in}

\section{Projected evolution operator} \label{par:Application of the protocol}
Both the field evolution and the survival probability, given by Eqs.\ (\ref{field density matrix evolution}) and (\ref{probability}), depend essentially on the projected evolution operator $\hat{V}_{\tau}$ defined in Eq.\ (\ref{contracted evolution operator}).
This operator, using Eq.\ (\ref{initial state}) as the state of the particle $\ket{\Phi_0}$ to be measured repeatedly,  results in
\begin{equation}\label{v2}
\hat{V}_{\tau}=\exp\!\left[-i \omega \tau\left( \hat{a}^{\dag}\hat{a}+\frac{1}{2}\right)\right]
\frac{ 1 }{\sqrt{2 \pi(\Delta p_{0})^2}} \int \mathrm{d} p\,\text{e}^{-f(p)},
\end{equation}
where
\begin{subequations}
\begin{align}
f(p)={}&\beta\frac{p^2}{2(\Delta p_{0})^2}\nonumber\\
&
{}-ig \frac{\sqrt{2(1-\cos\omega \tau) }}{\hbar \omega}( \hat{a}^{\dag}\mathrm{e}^{i \frac{\omega
     \tau}{2}}+\hat{a}\,\mathrm{e}^{-i\frac{\omega\tau}{2}})p,
\\
\beta
={}&1+ i\frac{(\Delta p_{0})^2\tau}{\hbar m}  \left(1 -\frac{2 mg^2}{ \hbar\omega }\frac{ \omega \tau-
  \sin \omega \tau}{\omega \tau}\right).
\end{align}
\end{subequations}
Introducing three independent dimensionless parameters
\begin{equation}\label{dimesionless parameters}
    \bar{\tau}=\omega \tau, \qquad \bar{g}=g \sqrt{\frac{m}{\hbar \omega}}, \qquad  \bar{\Delta}_p=\frac{\Delta p_{0}}{\sqrt{m\hbar \omega}},
\end{equation}
and performing the integration in Eq.\ (\ref{v2}) we obtain the projected evolution operator as
\begin{align}\label{v3}
    \hat{V}_{\tau}={}& M \exp\! \left[-i
\omega \tau
 \left( \hat{a}^{\dag}\hat{a}+\frac{1}{2}\right)\right] \nonumber \\ &\times \exp\!
\left[-\frac{G}{2} ( \hat{a}^{\dag}\mathrm{e}^{i  \frac{\omega
     \tau}{2}}+\hat{a}\,\mathrm{e}^{-i
     \frac{\omega
     \tau}{2}})^2 \right],
\end{align}
where
\begin{subequations}
\label{MFG}
\begin{equation}
M = \frac{1}{\sqrt{1+i\bar{\Delta}_p^2\bar{\tau}\left[1-2\bar{g}^2\left(1-\frac{\sin \bar{\tau}}{ \bar{\tau}}\right)\right]}},
\end{equation}
\begin{equation}
G  = 2M^2\bar{g}^2\bar{\Delta}_p^2(1-\cos \bar{\tau}).
\end{equation}
\end{subequations}

\vspace*{-0.1in}

\subsection{Diagonalization \label{par:Diagonalization}}
As already stated, in order to check if the conditions for purification are satisfied in the present setup we need to analyze the spectrum of $\hat{V}_{\tau}$. To diagonalize  the projected evolution operator $\hat{V}_{\tau}$ it is useful to rewrite its expression in Eq.\ (\ref{v3}) in a single unified exponential. The details are given in
Appendix \ref{app:Time-evolution operator}, where the projected evolution operator $\hat{V}_{\tau}$ is arranged in Eq.\ (\ref{eqn:coefficinets}) in the form
\begin{multline}\label{v4}
    \hat{V}_\tau=M\exp  \Biggl\{\frac{\ln(q-\sqrt{q^2-1})}{\sqrt{q^2-1}}G\\  \times\left[ \frac{\hat{a}^{\dag2}+ \hat{a}^2}{2} +\tilde{q} \left(\hat{a}^\dag \hat{a}+\frac{1}{2}\right) \right] \Biggr\},
\end{multline}
where
\begin{equation}\label{q e q}
 q=
\cos \bar{\tau}
+i G\sin\bar{\tau}, \qquad \tilde{q}= \cos  \bar{\tau} +\frac{i}{G}\sin  \bar{\tau}.
\end{equation}

It is diagonalized by the similarity transformation
\begin{equation}
\e^{\eta \hat{A}}\e^{\zeta \hat{A}^\dag}\hat{V}_\tau\e^{-\zeta \hat{A}^\dag}\e^{-\eta \hat{A}},
\end{equation}
where
\begin{equation}\label{xi e ni}
\zeta=\tilde{q} \pm \sqrt{ \tilde{q}^2-1} , \qquad \eta=\pm\frac{1}{2\tilde{q}},
\end{equation}
and then $\hat{V}_\tau$ is transformed to
\begin{equation}\label{V diagonalizzato}
    \hat{V}_\tau \to M \exp\! \left[\mp \left(\hat{a}^\dag \hat{a}+\frac{1}{2}\right) \ln(q-\sqrt{q^2-1})\right].
\end{equation}
This shows that the eigenvalues $\gamma_n $ and the right and left eigenstates, $\ket{u_n}$ and $\bra{v_n}$, of
\begin{equation}
\hat{V}_\tau =\sum_n \gamma_n \ket{u_n}\bra{v_n}
\end{equation}
are given by
\begin{subequations}
\label{eigenstates and eigenvalues}
\begin{gather}
   \gamma_n = M \exp\! \left[\mp  (n+1/2) \ln(q-\sqrt{q^2-1})\right],\\
   \ket{u_n}=\e^{-\zeta \hat{A}^\dag}\e^{-\eta \hat{A}}\ket{n},\quad
\bra{v_n}=\bra{n}\e^{\eta \hat{A}}\e^{\zeta \hat{A}^\dag},
\end{gather}
\end{subequations}
where $\ket{n}$ is the eigenstate of the number operator $\hat{a}^\dag\hat{a}$ belonging to its eigenvalue $n=0,1,\ldots$

A comment is in order as to the choice of the sign ($\pm$) in Eqs.\ (\ref{xi e ni}), (\ref{V diagonalizzato}), and (\ref{eigenstates and eigenvalues}).
As shown in Appendix \ref{app:Choise of sign}, in order to assure the normalizability of the eigenstates, for a given $q$, a correct sign must be chosen so that the inequality $|q \pm \sqrt{q^2-1}|=|q \mp \sqrt{q^2-1}|^{-1}<1$ holds.
The real part of $\ln(q-\sqrt{q^2-1})$ is equal to  $\ln\!|q-\sqrt{q^2-1}|=-\ln\!|q+\sqrt{q^2-1}|$ and it is easy to verify that one has to choose the upper(lower) sign, that is, $+(-)$ in Eq.~(\ref{xi e ni}) and $-(+)$ in Eqs.~(\ref{V diagonalizzato}) and (\ref{eigenstates and eigenvalues}) if the real part is positive(negative).
Therefore, $\pm$ may be substituted with the sign of the real part of $\ln(q-\sqrt{q^2-1})$.
We observe that when $\bar{\tau}=(2\ell+1)\pi$ ($\ell=0,1,\ldots$) we have $q=0$ and then $|q \pm \sqrt{q^2-1}|=1$, independently of the choice of sign, so that for these values of $\bar{\tau}$ the normalizability of the eigenstates is lost and there is no distillation.

\vspace*{-0.1in}

\section{Distillation \label{par:Distillation limit}}
Analyzing the structure of the eigenvalues $\gamma_n $ of $\hat{V}_{\tau}$ in Eq.\ (\ref{eigenstates and eigenvalues}) we see that the largest (in magnitude) eigenvalue $\gamma_0$ is unique, discrete, and nondegenerate, so that in the large $N$ limit, the operator $\hat{V}_{\tau}^N$ is dominated by a single term \cite{Nakazato2003}
\begin{equation}\label{dominant term}
    \hat{V}_{\tau}^N \xrightarrow{\text{large } N}\gamma_0^N \ket{u_0}\bra{v_0}.
\end{equation}
Then, using Eq.\ (\ref{dominant term}) in Eq.\ (\ref{field density matrix evolution}), in the large $N$ limit for a nonvanishing $\tau$, the state of the field asymptotically approaches the pure state,
\begin{equation}\label{stato distillato}
     \hat{\rho}_F^{\tau}(N)\xrightarrow{\text{large } N} \frac{\ket{u_0}\bra{u_0}}{\langle u_0|u_0\rangle}=\ket{\xi}\bra{\xi}.
\end{equation}
Notice that the pure state $\ket{u_0}$ is explicitly written as
\begin{equation}
\ket{u_0}=\e^{-\zeta \hat{A}^\dag}\ket{0}=\sqrt{\cosh {r}}\,\hat{S}(\xi)\ket{0}=\sqrt{\cosh {r}}\ket{\xi},
\end{equation}
where $\hat{S}(\xi)=\exp( -\frac{\xi}{2}\hat{a}^{\dag 2}-\frac{\xi^*}{2}\hat{a}^{2})$ is a squeezing operator with $ \xi= {r} \e^{i\varphi}$.
The squeezing parameter $r$ and the phase $\varphi$ are given by
\begin{equation}\label{xi distilled}
 r=\tanh^{-1}|  \zeta|,
\qquad
\varphi= \arg\zeta,
\end{equation}
where $\zeta $ is defined in Eq.\ (\ref{xi e ni}) and $\tilde{q}$ in Eq.\ (\ref{q e q}).
Therefore, Eq.\ (\ref{stato distillato}) shows that the field mode is distilled into a squeezed state.
The final pure squeezed state $\ket{\xi}$ is independent of the choice of the initial state of the field, i.e., any initial (eventually mixed) state shall be driven to the unique pure state $\ket{\xi}$ by the repeated measurements performed on the particle to confirm it to be in the Gaussian state $\ket{\Phi_0}$.

There are four, the first two independent of and the last two dependent on the initial state of the field mode, relevant quantities characterizing this distillation process:

\begin{itemize}
  \item Speed of distillation: the purification is achieved quickly if $|\gamma_n/\gamma_0|=|\gamma_1/\gamma_0|^n
 \ll 1$ for $n\neq 0$.
This means that we have a quick distillation if
 \begin{equation}\label{quick distillation}
    |\gamma_1/\gamma_0|=\exp\!\left[-\left|\ln\bigl|q-\sqrt{q^2-1}\bigr|\right|\right]\ll 1.
\end{equation}
The quickness of distillation is thus linked to $\left|\ln\bigl|q-\sqrt{q^2-1}\bigr|\right|=\left|\ln\bigl|q+\sqrt{q^2-1}\bigr|\right|$.
  \item Degree of squeezing of the distilled state $\ket{\xi}$: this is given by $r$ in Eq.\ (\ref{xi distilled}). It regulates the average number of quanta $\langle\hat{a}^{\dag}\hat{a}\rangle=\sinh^2r$ and its variance $\langle(\hat{a}^\dag\hat{a})^2\rangle-\langle\hat{a}^\dag\hat{a}\rangle^2=2 \sinh^2r\,(\sinh^2r +1)$.
For a given $r$, $\varphi$ determines the variance  $\Delta_{\hat{x}_\vartheta}^2=
\langle(\hat{x}_\vartheta)^2\rangle-
\langle \hat{x}_\vartheta \rangle^2$ of the quadrature operator $\hat{x}_\vartheta =\frac{1}{\sqrt{2}}(\hat{a}\,\e^{-i \vartheta} + \hat{a}^{\dag}\e^{i \vartheta})$, where $\vartheta$ is a real phase.
Observe that $\e^{-2 r}/2 < \Delta_{\hat{x}_\vartheta}^2 <\e^{2r}/2$.
If $\sin^2(\vartheta-\varphi) < (\e^{2r}+1)^{-1}$ the variance $\Delta_{\hat{x}_\vartheta}^2$ is less than $1/2$ and in this case the quadrature $\hat{x}_\vartheta$ is squeezed.
  \item Probability of success of the protocol: this is represented by the survival probability $P_{\tau}(N)$ introduced in Eq.\ (\ref{probability}).
It is desirable to have a higher probability of success to keep higher yields when the distillation has been attained.
While the state approaches the squeezed state $\ket{\xi}$ as shown in (\ref{stato distillato}), the probability behaves as
\begin{equation}
P_\tau(N)
\xrightarrow{\text{large }N}|\gamma_0|^{2N}\bracket{u_0}{u_0}\bra{v_0}\hat{\rho}_F(0)\ket{v_0}.
\label{eqn:DecayP}
\end{equation}
Therefore, it is preferable to have a larger $|\gamma_0|$ (closer to unity) for a slower decay of the probability $P_\tau(N)$.
  \item Fidelity: it indicates how close the extracted state of the field mode is to the target pure state after $N$ measurements.
  For an efficient distillation protocol the state after $N$ measurements must be as close to the final target state as possible.
The fidelity is defined by
\begin{equation}\label{fidelity}
    F_\tau(N)=\langle \xi| \hat{\rho}_F^{\tau}(N)|\xi \rangle,
\end{equation}
which is a positive number and approaches unity as the field state $\hat{\rho}_F^{\tau}(N)$ becomes close to the target pure state $|\xi\rangle$.
\end{itemize}

In the following we shall analyze the dependence of these quantities on the parameters $\bar{\tau}$, $\bar{g}$, and $\bar{\Delta}_p$.
Our aim is to find  optimal values of the parameters satisfying two independent requirements:
\begin{enumerate}
  \item fast distillation with high degree of squeezing [see (\ref{quick distillation})], and
  \item high fidelity with a sufficiently high probability of success of the protocol [see (\ref{eqn:DecayP})].
\end{enumerate}

Even if the second condition is fundamental for an efficient protocol for the distillation,
we start discussing the first condition, concerning  the speed of distillation and the degree of squeezing, because it involves quantities which are independent of the initial state of the field mode.
\vspace*{-0.1in}

\subsection{Distillation speed vs squeezing \label{par:Distillation rapidity vs Squeezing}}
\begin{figure}
%\begin{center}
\includegraphics[width=7.2 cm, height=6.5 cm]{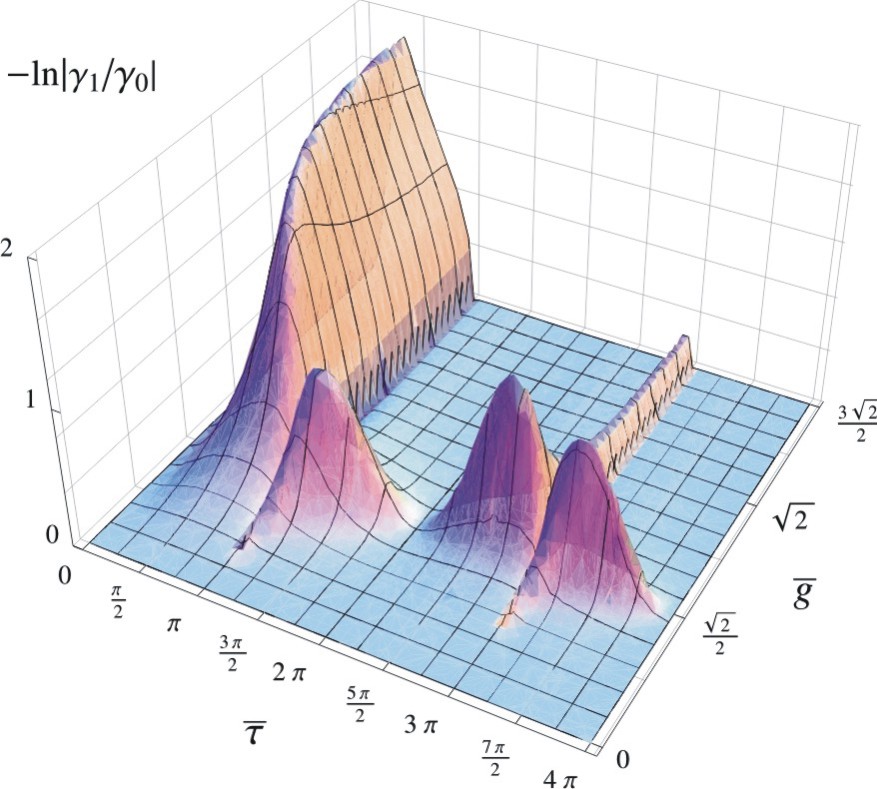}
\caption{\label{FigReK_tg}The ratio between the first two eigenvalues $-\ln|\gamma_1/\gamma_0|$ as a function of the dimensionless parameters $\bar{\tau}$ and $ \bar{g}$ with fixed $  \bar{\Delta}_p=1$.}
%\end{center}
\end{figure}

In order to have indications for the ranges of the parameters where there is a quick distillation, we plot the ratio between the first two eigenvalues $-\ln|\gamma_1/\gamma_0|=\Bigl|\ln\bigl|q\mp\sqrt{q^2-1}\bigr|\Bigr|$, looking for regions where this quantity becomes large.
Figure \ref{FigReK_tg} shows $-\ln|\gamma_1/\gamma_0|$ as a function of the dimensionless parameters in Eq.\ (\ref{dimesionless parameters}), $\bar{\tau}$ and $ \bar{g}$ with fixed $  \bar{\Delta}_p=1$.
The plot evidences that, for a fixed $  \bar{\Delta}_p$, the ratio $-\ln|\gamma_1/\gamma_0|$ has a strong dependence on the values of $\bar{g}$ and $\bar\omega$.
In particular, for $\bar{\tau} \lesssim \pi $ and $\bar{g}\gtrsim1/\sqrt{2}$, the ratio can be greater than unity and in this region a fast distillation is available.

\begin{figure}
%\begin{center}
\includegraphics[width=7.2 cm, height=6.5 cm]{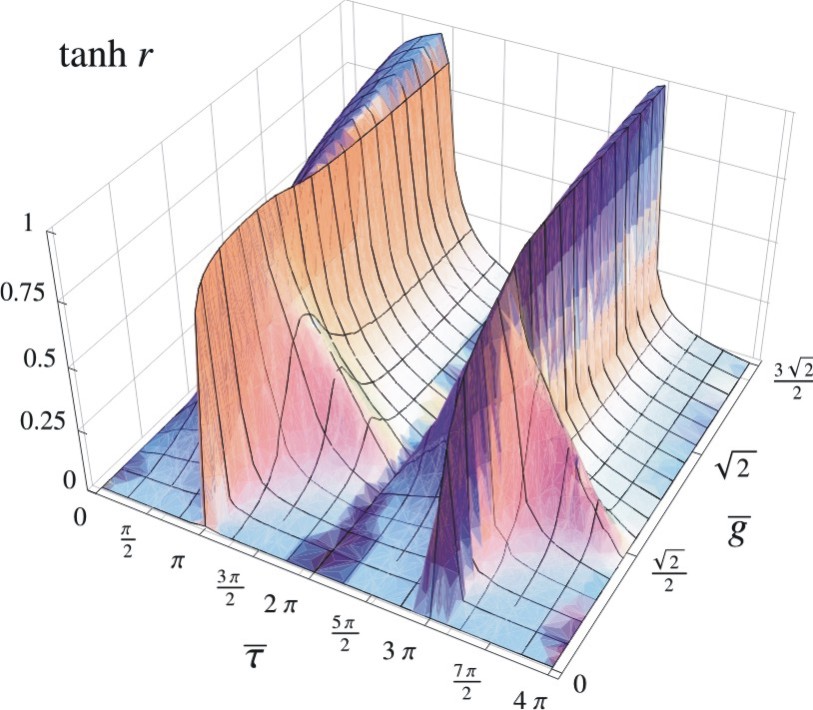}
\caption{\label{Figr_tg}$\tanh r$ as a function of the dimensionless parameters $\bar{\tau}$ and $\bar{g}$ with fixed $  \bar{\Delta}_p=1$.}
%\end{center}
\end{figure}

Next, to investigate the degree of squeezing of the distilled state,
we plot the hyperbolic tangent of the squeezing parameter, $\tanh r=|\zeta|$,
instead of the squeezing parameter $r$ itself, in order to avoid divergences in ${r}$ when $|\zeta|\to1$.
Figure \ref{Figr_tg} shows $\tanh r$ as a function of the dimensionless parameters $\bar{\tau}$ and $ \bar{g}$ with fixed $  \bar{\Delta}_p=1$.
It is evident that for odd multiples of $\pi$  (values of $\tau$ not allowed for distillation) and $\bar{g}\neq 0$ we have $\tanh r \rightarrow 1$ and thus $r\rightarrow \infty $. With $\bar{\tau}$ just smaller than these values the distilled state is highly squeezed and in particular
the region $\bar{\tau}\lesssim \pi$ and $\bar{g}\gtrsim1/\sqrt{2}$ appears to be more appropriate to obtain states with a high degree of squeezing.
For  $ \bar{g}$ and $\bar{\Delta}_p$ large enough, $\tanh r$ depends only on $\tau$.

\begin{figure}
%\begin{center}
\includegraphics[width=7.2 cm, height=4.5 cm]{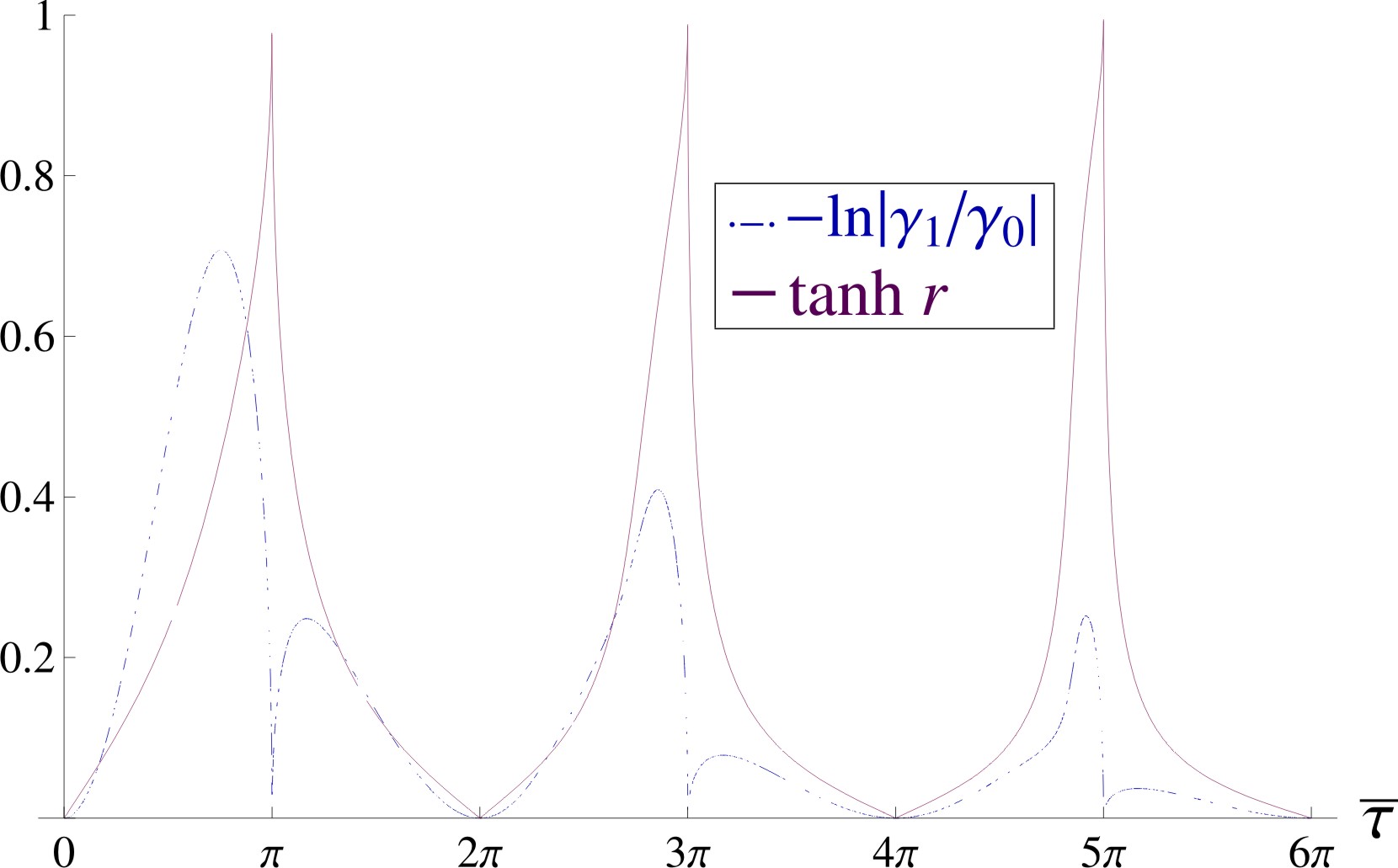}
\caption{\label{Figkr_t} $-\ln|\gamma_1/\gamma_0|$ (dotdashed line) and  $\tanh r$ (solid line), as functions of the dimensionless parameter $\bar{\tau}$ with $ \bar{g}=1$ and $\bar{\Delta}_p=0.4$.}
%\end{center}
\end{figure}

The above plots indicate that distillation speed, linked to $-\ln|\gamma_1/\gamma_0|$, and the degree of squeezing, linked to $\tanh r$, have strong and different dependencies on the parameters. Now we show that it is possible to find values of the  parameters where quick distillation and strong squeezing would be attainable simultaneously.
In Fig.\ \ref{Figkr_t} we compare
the behaviors of $-\ln|\gamma_1/\gamma_0|$ (dotdashed line) and of $\tanh r$ (solid line), as functions of the dimensionless parameter $\bar{\tau}$ with   $ \bar{g}=1$ and $\bar{\Delta}_p=0.4$.
From the plot one sees that  $-\ln|\gamma_1/\gamma_0|\gtrsim0.5 $ (sufficiently quick distillation) and $\tanh r \gtrsim0.5$ (sufficiently strong squeezing) can be fulfilled for $\pi/2\lesssim\omega \tau\lesssim\pi$.

\vspace*{-0.1in}

\subsection{Survival probability vs fidelity \label{par:Survival probability vs Fidelity}}
Both the survival probability and the fidelity depend on the initial state of the field. We consider the case where the field is initially in a coherent state $\ket{\alpha}$. Although the survival probability and the fidelity can be analytically obtained when the field is initially in a coherent state, their explicit expressions are rather involved so that in the following we will just present the results.

\begin{figure}
%\begin{center}
\includegraphics[width=7.2 cm, height=4.75 cm]{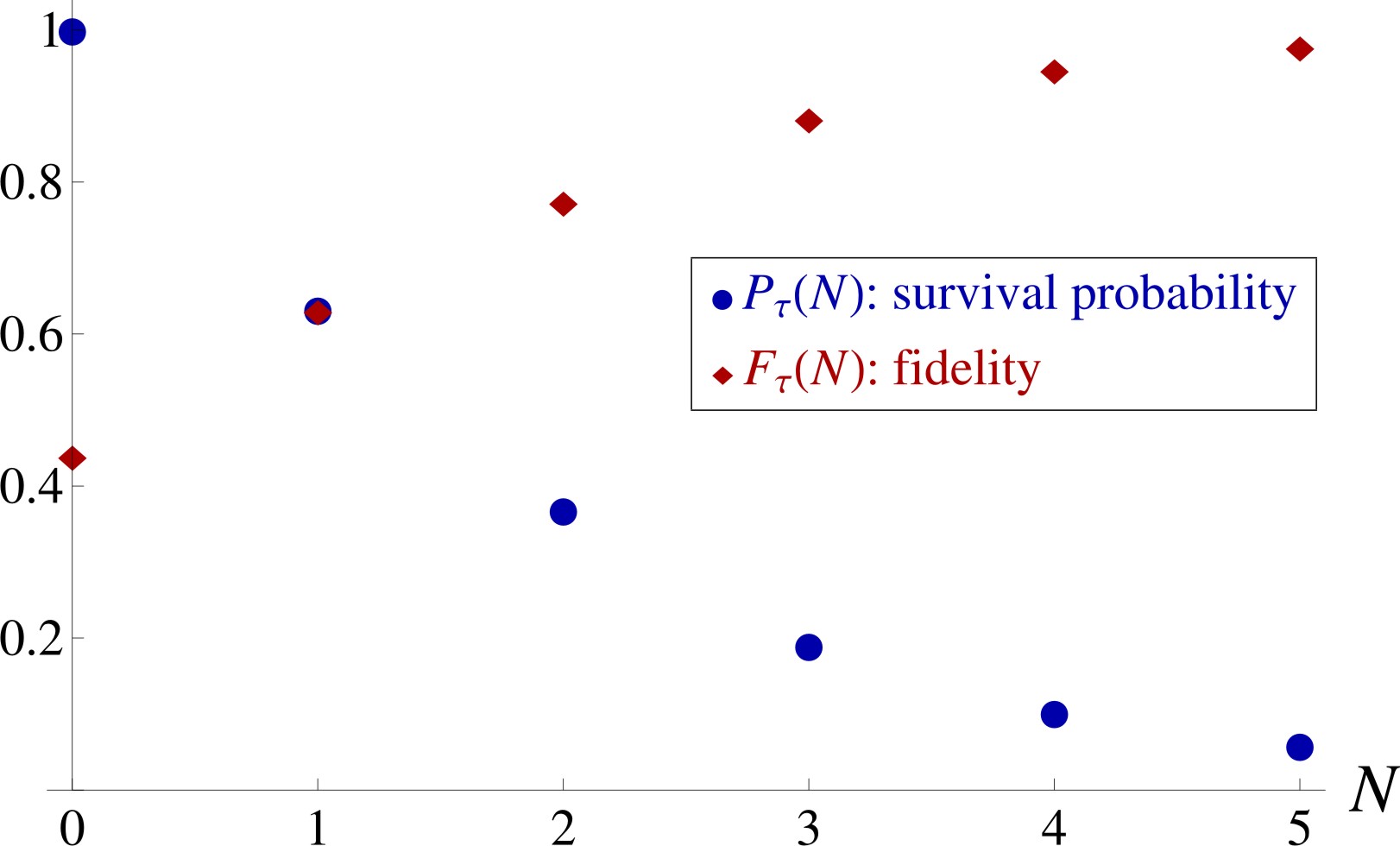}
\caption{\label{probabilitavsfidelity}$P_{\tau}(N)$ (circles) vs $ F_{\tau}(N)$ (diamonds) as functions of the number of measurements $N$ with $ \bar{g}=1$, $  \bar{\Delta}_p=0.4$, and $\bar{\tau}= 0.9\pi$.
The field is initially in a coherent state $\ket{\alpha}$ of amplitude $\alpha=1$.}
%\end{center}
\end{figure}

Now, in order to check if the present protocol gives distillation with a good probability of success in this case, we compare the evolutions of the survival probability and of the fidelity for a given $\tau$. In particular we are looking for values of the parameters such that when the fidelity gets close to 1 the survival probability is still high enough.
In Fig.\ \ref{probabilitavsfidelity} the evolutions of the survival probability and of the fidelity for $\alpha=1$ and $\bar{\tau}= 0.9\pi$ are given. The values of the other parameters are the same as in Fig.\ \ref{Figkr_t}.
From the plot one sees that high values of fidelity can be obtained with the survival probability still far from 0.

These two plots, Figs.\ \ref{Figkr_t} and \ref{probabilitavsfidelity}, clearly indicate that our protocol allows one to generate efficiently, that is, with a high fidelity and a
finite (nonvanishing) probability, pure squeezed states with a
sufficiently high degree of squeezing.  Indeed, for example, if we tune
the parameters as $\bar{g}=1$, $\bar{\Delta}_p=0.4$ and perform
measurements with a period of $\bar\tau=0.9\pi$, we will obtain a well
squeezed state with a squeezing parameter
$r\sim\tanh^{-1}0.6\sim0.54$ with a fidelity of
$\sim80\%$ and a probability $\sim20\%$ already after a couple of
measurements $N\sim2,3$.

\vspace*{-0.1in}
\section{Conclusions \label{par:Conclusion}}
In general for a bipartite system made of two interacting parts it is known that Zeno-like measurements on a part may lead, under certain conditions, the non-measured part towards a pure state independently of its initial configuration. The general procedure has been so far analyzed in the case where the measured part has a discrete spectrum, while  it is not obvious whether the distillation can be obtained when the measured system has a continuous spectrum.
Here we have analyzed this topic considering a specific bipartite system consisting of a particle, characterized by a continuous spectrum, interacting with a field mode characterized by a discrete spectrum but with an infinite number of levels. The present distillation protocol consists of  repeatedly  projecting the particle to a Gaussian state. The projected evolution operator that regulates the field-mode dynamics between the consecutive measurements performed on the particle has been obtained. It has been shown that, with the measurement protocol chosen, the spectrum of this operator is discrete and satisfies the criteria that allow one to obtain a field-mode distillation. As a consequence of the protocol, the field is driven to a squeezed state independently of its initial state.
The dependencies of the distillation speed, that is connected to the ratio between the first two eigenvalues of the projected evolution operator, and of the characteristics of the distilled state, i.e., the squeezing parameter, are investigated as functions of parameters such as the interval between two measurements, the particle-field mode coupling constant, and the width of the particle's Gaussian state.
Varying the values of the parameters different regimes are observed and we have shown that it is possible to choose values such that one has both quick distillation and strong squeezing and/or high values of fidelity and finite, well far from zero, values of probability of success of the measurement protocol.

\vspace*{-0.1in}
\acknowledgments
This work is supported by the bilateral Italian-Japanese Projects II04C1AF4E on ``Quantum Information, Computation and Communication'' of the Italian Ministry of Education, University and Research, by the Joint Italian-Japanese Laboratory on ``Quantum Information and Computation'' of the Italian Ministry for Foreign Affairs, by the
Grant-in-Aid for Scientific Research (C) from the Japan Society for
the Promotion of Science, and by a Special Coordination Fund for Promoting Science and Technology and the Grant-in-Aid for Young Scientists (B) (No.\ 21740294) both from the Ministry of Education, Culture, Sports, Science and Technology, Japan.

\vspace*{-0.2in}

\appendix
\section{Unification of exponential factors in $V_\tau$}\label{app:Time-evolution operator}
The projected time-evolution operator $\hat{V}_\tau$ of Eq.\ (\ref{v3}) is of the following form:
\begin{equation}
\hat{V}_\tau
=M\e^{-i\omega\tau \hat{B}}\e^{-G(\e^{i\omega \tau}\hat{A}^\dag+\e^{-i\omega \tau}\hat{A}+\hat{B})},
\label{eqn:V}
\end{equation}
where
\begin{equation} \label{capital operators}
\hat{A}=\frac{1}{2}\hat{a}^2,\qquad
\hat{B}=\hat{a}^\dag \hat{a}+\frac{1}{2}.
\end{equation}
Note the Lee algebra among $\hat{A}$, $\hat{A}^\dag$, and $\hat{B}$,
\begin{equation} \label{capital operators commutators}
[\hat{A},\hat{A}^\dag]=\hat{B},\qquad
[\hat{A},\hat{B}]=2\hat{A},\qquad
[\hat{A}^\dag,\hat{B}]=-2\hat{A}^\dag.
\end{equation}
For these generators of the algebra, the following formula for the factorization of exponential is available:{
\begin{equation}
\e^{\mu  \hat{A}^\dag+\nu  \hat{A}+\lambda  \hat{B}}
=\e^{x\hat{A}^\dag}\e^{y\hat{B}}\e^{z \hat{A}},
\label{eqn:Factorize}
\end{equation}
where
\begin{equation}
\begin{cases}
\medskip
\displaystyle
x=\frac{(\mu /\kappa)\tanh\kappa}{
1-(\lambda /\kappa)\tanh\kappa
},\\
\medskip
\displaystyle
y=-\frac{1}{2}\ln\!\left(
\cosh\kappa-\lambda \frac{\sinh\kappa}{\kappa}
\right)^2,\\
\displaystyle
z=\frac{(\nu /\kappa)\tanh\kappa}{
1-(\lambda /\kappa)\tanh\kappa
},
\end{cases}
\kappa=\sqrt{\lambda ^2-\mu \nu }.
\label{eqn:Coefficients}
\end{equation}
}
Inverse relations are given by
\begin{equation}
\begin{cases}
\medskip
\displaystyle
\mu =\frac{\kappa}{\sinh\kappa}x\e^{-y},\\
\medskip
\displaystyle
\nu =\frac{\kappa}{\sinh\kappa}z\e^{-y},\\
\displaystyle
\lambda =-\sqrt{\kappa^2+\mu \nu },
\end{cases}
\kappa
=\cosh^{-1}\!
\frac{1}{2}(
\e^{y}
+\e^{-y}
-x z\e^{-y}
).
\label{eqn:Inversion}
\end{equation}
Note the formula for the reciprocal function
\begin{equation}
\cosh^{-1}\!x
=\pm\ln(x+\sqrt{x^2-1}).
\end{equation}

Now, let us come back to the projected time-evolution operator (\ref{eqn:V}).
By making use of the factorization formula (\ref{eqn:Factorize})--(\ref{eqn:Coefficients}), one has
\begin{equation}
\hat{V}_\tau
=M\e^{-i\omega\tau \hat{B}}
\e^{-\frac{G}{1+G}\e^{i \omega
     \tau}\hat{A}^\dag}
\e^{-\ln(1+G)\hat{B}}
\e^{-\frac{G}{1+G}\e^{-i\omega \tau}\hat{A}}.
\end{equation}
By exchanging the order of the first two exponentials, one obtains
\begin{equation} \label{eqn:Unified}
\hat{V}_\tau
=M
\e^{-x\hat{A}^\dag}
\e^{-y\hat{B}}
\e^{-x\hat{A}},
\qquad \begin{cases}
\medskip
\displaystyle
x=\frac{G}{1+G}\e^{-i \omega
     \tau},\\
\displaystyle
y=\ln(1+G)+i\omega\tau .
\end{cases}
\end{equation}
Then, we unify the exponentials via the formula (\ref{eqn:Factorize}) with (\ref{eqn:Inversion}) to obtain
\begin{align} \label{eqn:coefficinets}
&\hat{V}_\tau=M\e^{\mu  \left(\hat{A}^\dag+ \hat{A}\right)+\lambda  \hat{B}}, \nonumber
\displaybreak[0]\\
& \begin{cases}
\medskip
\displaystyle
\mu =\frac{G}{\sqrt{q^2-1}}\ln(q-\sqrt{q^2-1})
 \,\\
\displaystyle
\lambda =\frac{G\cos\omega\tau+i\sin\omega\tau}{\sqrt{q^2-1}}\ln(q-\sqrt{q^2-1}),
\end{cases}
\end{align}
where $q$ is defined in (\ref{q e q}).
The expression (\ref{v4}) is thus obtained.

\vspace*{-0.2in}
\section{Choice of sign in diagonalization}\label{app:Choise of sign}
Equations (\ref{xi e ni}), (\ref{V diagonalizzato}), and (\ref{eigenstates and eigenvalues}) seem to show that the diagonalization procedure presented in the text admits two possible signs $\pm$.
The immediate concern, in such a case, would be whether the unitarity of the time-evolution operator $\hat{U}(\tau)$, which dictates that the absolute values of eigenvalues $\hat{V}_\tau$ are
strictly upper-bounded by unity, is preserved by the solution in Eq.\ (\ref{eigenstates and eigenvalues}), or which would be the right choice of the signs
$(\mp)$ of the eigenvalues (and the eigenstates) if only one of them can be allowed.
It will be shown here that the normalizable eigenstates are those where $|q -\sqrt{q^2 - 1}| < 1$, i.e., belonging to the eigenvalues with their magnitudes
always less than unity,
\begin{equation}\label{dim0}
    \left|\gamma_n\right| \le\exp\! \left[  (n+1/2) \ln\bigl| q\pm\sqrt{q^2-1}\bigr|\right]<1.
\end{equation}
The normalizability of the eigenstate $\ket{u_0}$ for  $n = 0$,
\begin{equation}\label{dim1}
    \| \e^{-\zeta \hat{A}^\dag}\e^{-\eta \hat{A}}\ket{0}\|<\infty,
\end{equation}
is sufficient to show the above statement.
The left-hand side is calculated to be
\begin{align}\label{dim2}
    \bra{0}\e^{-\zeta^*\hat{A}}\e^{-\zeta\hat{A}^\dag}\ket{0}
&=\int \frac{\text{d}^2\alpha}{\pi}\,
    \e^{-\frac{\zeta^*}{2} \alpha^2}\e^{-\frac{\zeta}{2} (\alpha^*)^2}|\bra{0}\alpha \rangle|^2 \nonumber
\displaybreak[0]
\\
&=\int\frac{\text{d}\alpha_\text{R}\,\text{d}\alpha_\text{I}}{\pi}\,\e^{-\alpha \cdot \mathcal{A} \alpha},
\end{align}
where the exponent is explicitly written as
\begin{equation}\label{dim3}
    \alpha \cdot \mathcal{A} \alpha=
\begin{pmatrix}\alpha_\text{R}&\alpha_\text{I}
\end{pmatrix}\begin{pmatrix}
1+\zeta_\text{R} & \zeta_\text{I} \\
\zeta_\text{I} & 1-\zeta_\text{R}
\end{pmatrix}\begin{pmatrix}
\alpha_\text{R} \\
\alpha_\text{I}
\end{pmatrix},
\end{equation}
and $\alpha_\text{R}=\Re\alpha$, $\alpha_\text{I}=\Im\alpha$, $\zeta_\text{R}=\Re\zeta$, $\zeta_\text{I}=\Im\zeta$.
The eigenvalues of the matrix $\mathcal{A}$ are easily found to be $1 \pm |\zeta|$, both of which have to be positive, i.e.,
$
    |\zeta|<1,
$
in order for the above state is normalizable. That is, the normalizability of the eigenstates is assured if the condition
$|\zeta| < 1$ is satisfied. This condition is explicitly written as that for $q$ and $G$
\begin{equation}\label{dim5}
   | \zeta|=\bigl|\tilde{q}\pm \sqrt{\tilde{q}^2-1}\bigr|<1,
\end{equation}
which just reduces to
\begin{equation}\label{dim6}
    \bigl|q \pm \sqrt{q^2-1}\bigr|<1\,,
\end{equation}
if $G$ is replace with $G^{-1}$. The normalizability condition of the eigenstates for the case of $G^{-1}$ thus ensures the unitarity in the case of $G$. Stated differently, we have to choose an appropriate sign between $+$ and $-$ so that the absolute
value of the argument of the logarithm satisfies the above inequality in order for the eigenstates to be normalizable. The
unitarity is always satisfied, or we just have to make an appropriate choice of the phase of the square root $\sqrt{q^2 -1} =
\sqrt{(\cos \omega \tau + iG \sin \omega \tau)^2 - 1}$, to which the sign $(\pm)$ could be considered to be absorbed.

\end{document}